%% file: main.tex
\newif\ifanon

\ifanon
    \documentclass[manuscript,review,anonymous]{acmart}
    \newcommand{\school}{\textbf{<Redacted>}\xspace}
    \newcommand{\learningEnvironment}{\textbf{<Redacted>}\xspace}
\else
    \documentclass[manuscript]{acmart}
    \newcommand{\school}{Vanderbilt University\xspace}
    \newcommand{\learningEnvironment}{RedForest\xspace}
\fi
\AtBeginDocument{%
  }

\setcopyright{acmlicensed}
\copyrightyear{2025}
\acmYear{2025}
\acmDOI{XXXXXXX.XXXXXXX}
\acmConference[IUI '26]{Proceedings of the 31st International Conference on Intelligent User Interfaces}{March 23--26, 2026}{Paphos, Cyprus}
\acmISBN{978-1-4503-XXXX-X/2018/06}

\usepackage{subfig}
\usepackage{tabularx}
\usepackage{tabto}
\usepackage{makecell}
\usepackage{xspace}

\begin{document}

\title{Designing Gaze Analytics for ELA Instruction: A User-Centered Dashboard with Conversational AI Support}

\author{Eduardo Davalos}
\email{davalosedu515@trinity.edu}
\orcid{0000-0001-7190-7273}
\affiliation{%
  \institution{Trinity University}
  \city{San Antonio}
  \state{TX}
  \country{USA}
}

\author{Yike Zhang}
\email{yzhang5@vanderbilt.edu}
\orcid{0000-0003-3503-2996}
\affiliation{%
  \institution{St. Mary's University}
  \city{San Antonio}
  \state{TX}
  \country{USA}
}

\author{Shruti Jain}
\email{shruti.jain@vanderbilt.edu}
\orcid{0009-0000-7853-0560}
\affiliation{%
  \institution{Vanderbilt University}
  \city{Nashville}
  \state{TN}
  \country{USA}
}

\author{Namrata Srivastava}
\email{namrata.srivastava@vanderbilt.edu}
\orcid{0000-0003-4194-318X}
\affiliation{%
  \institution{Vanderbilt University}
  \city{Nashville}
  \state{TN}
  \country{USA}
}

\author{Trieu Truong}
\email{trieu.vy.truong@vanderbilt.edu}
\orcid{0009-0009-7723-7069}
\affiliation{%
  \institution{Vanderbilt University}
  \city{Nashville}
  \state{TN}
  \country{USA}
}

\author{Nafees-ul Haque}
\email{nafees-ul.haque@vanderbilt.edu}
\orcid{0009-0006-5978-4227}
\affiliation{%
  \institution{Vanderbilt University}
  \city{Nashville}
  \state{TN}
  \country{USA}
}

\author{Tristan Van}
\email{tristan.v.van@vanderbilt.edu}
\orcid{0009-0007-0816-1879}
\affiliation{%
  \institution{Vanderbilt University}
  \city{Nashville}
  \state{TN}
  \country{USA}
}

\author{Jorge A. Salas}
\email{jorge.a.salas@vanderbilt.edu}
\orcid{0000-0002-8885-0813}
\affiliation{%
  \institution{Vanderbilt University}
  \city{Nashville}
  \state{TN}
  \country{USA}
}

\author{Sara McFadden}
\email{sara.mcfadden@vanderbilt.edu}
\orcid{0000-0001-6375-1272}
\affiliation{%
  \institution{Vanderbilt University}
  \city{Nashville}
  \state{TN}
  \country{USA}
}

\author{Sun-Joo Cho}
\email{sj.cho@vanderbilt.edu}
\orcid{0000-0002-2600-8305}
\affiliation{%
  \institution{Vanderbilt University}
  \city{Nashville}
  \state{TN}
  \country{USA}
}

\author{Gautam Biswas}
\email{gautam.biswas@vanderbilt.edu}
\orcid{0000-0002-2752-3878}
\affiliation{%
  \institution{Vanderbilt University}
  \city{Nashville}
  \state{TN}
  \country{USA}
}

\author{Amanda Goodwin}
\email{amanda.goodwin@vanderbilt.edu}
\orcid{0000-0002-6439-7399}
\affiliation{%
  \institution{Vanderbilt University}
  \city{Nashville}
  \state{TN}
  \country{USA}
}

\renewcommand{\shortauthors}{Davalos et al.}

\begin{abstract}
Eye-tracking offers rich insights into student cognition and engagement, but remains underutilized in classroom-facing educational technology due to challenges in data interpretation and accessibility. In this paper, we present the iterative design and evaluation of a gaze-based learning analytics dashboard for English Language Arts (ELA), developed through five studies involving teachers and students. Guided by user-centered design and data storytelling principles, we explored how gaze data can support reflection, formative assessment, and instructional decision-making. Our findings demonstrate that gaze analytics can be approachable and pedagogically valuable when supported by familiar visualizations, layered explanations, and narrative scaffolds. We further show how a conversational agent, powered by a large language model (LLM), can lower cognitive barriers to interpreting gaze data by enabling natural language interactions with multimodal learning analytics. We conclude with design implications for future EdTech systems that aim to integrate novel data modalities in classroom contexts.
\end{abstract}




\maketitle

\section{Introduction}

Educational technologies (EdTech) are rapidly transforming the way students learn, offering new opportunities to personalize instruction \cite{azevedo_lessons_2022} and enhance engagement \cite{psaltis_multimodal_2018}. However, a persistent limitation of many modern EdTech applications is their exclusive reliance on log data, such as clickstreams, time-on-task, or quiz attempts, to interpret and analyze learner behavior \cite{worsley_multimodal_2018}. While these digital traces offer some insight, they fail to capture the full richness of how students engage cognitively and behaviorally with learning materials \cite{blikstein_multimodal_nodate}.

Recent advances in multimodal learning analytics (MMLA) have pushed the field toward more comprehensive, sensor-informed understandings of student behavior \cite{SchneiderSteppingInsights, ochoa_chapter_nodate}. Among these modalities, eye gaze stands out as particularly critical for human-computer interaction and user interface design. Gaze offers a window into cognitive processes such as information acquisition \cite{Srivastava2021AreTracking}, reading strategies \cite{Chen2023CharacteristicsStudy}, and attention distribution \cite{emerson_early_2020}, factors that are especially important in English Language Arts (ELA) learning tasks like silent reading comprehension \cite{Just1980AComprehension,Pena2024EyeTracking, Meziere2023UsingComprehension, Meziere2024ScanpathComprehension}, where student engagement is largely invisible to teachers.

Despite its promise, gaze data in educational research often stops at academic analysis: gaze patterns are collected, analyzed, and published, but rarely are the findings translated into actionable insights for teachers or students \cite{knoop-van_campen_how_2021}. This gap between research and practice is narrowing, however, with the emergence of affordable, webcam-based eye-tracking technologies \cite{heck_webcam_2023, papoutsaki_webgazer_2016}, these tools make it increasingly feasible to integrate real-time gaze analytics into classroom-facing applications. Yet, a major hurdle remains: how to communicate gaze data effectively to users who are unfamiliar with it -- namely, teachers and students.

Gaze is a novel and abstract data modality for most educators. Interpreting gaze heatmaps, fixation sequences, or scanpaths requires expertise in eye-tracking that few teachers possess or trained on \cite{knoop-van_campen_how_2021}. This raises critical design challenges around accessibility, usability, and pedagogical utility. Moreover, traditional dashboards may not suffice for helping users make sense of such unfamiliar data.

\subsection{Research Questions}
To address these challenges, we explore how data storytelling and conversational interfaces can enhance the interpretability and pedagogical relevance of gaze-based learning analytics. Specifically, we investigate the following research questions:
\begin{itemize}
    \item \textbf{RQ1}: Would teachers and students find gaze-based analytics approachable and capable of offering actionable insights?
    \item \textbf{RQ2}: How can we apply data storytelling principles to the design of gaze analytics tools for ELA instruction?
    \item \textbf{RQ3}: Could the use of a conversational agent assist in the navigation and interpretation of gaze data within an analytics dashboard?
\end{itemize}

By investigating these questions, this work aims to bridge the gap between the technical complexity of gaze analytics and the practical needs of classroom stakeholders, advancing the development of intelligent user interfaces that make cutting-edge learning data both accessible and impactful.

\section{Background}

\subsection{Data Storytelling in Learning Analytics}
Developing technology to support complex human tasks, particularly in education and learning environments, requires active stakeholder involvement to ensure that the system aligns with user needs and enhances real-world workflows \cite{sarmiento_participatory_2022}. In educational technology, where systems are designed to support teachers, students, and researchers, a user-centered design (UCD) methodology provides a structured approach to incorporating user feedback at every stage of development. UCD has been widely used in the development of EdTech systems, including learning analytics dashboards that present student performance data in an interpretable manner \cite{alzoubi_concept_2024}, and data visualizations that support instructional decision-making by making complex insights more actionable \cite{echeverria_driving_2018}. These techniques contribute to the broader field of data storytelling, where well-designed representations of student engagement, learning behaviors, and academic progress enable educators to make informed pedagogical choices \cite{fernandez-nieto_storytelling_2021,pozdniakov_how_2023}. By integrating stakeholder perspectives throughout the design process, UCD ensures that educational technology is functional, intuitive, accessible, and impactful in real-world learning environments.

UCD in educational technology goes beyond usability and functionality; it plays a key role in shaping how data is presented, interpreted, and acted upon. A growing body of research highlights the importance of data storytelling in learning analytics \cite{martinez-maldonado_data_2020}, where complex multimodal data must be translated into actionable insights for teachers and students. Research has shown that effective dashboard design must align with teachers' learning goals, enabling them to derive meaningful narratives from student data rather than simply navigating static visualizations \cite{echeverria_exploratory_2018}. However, many existing EdTech solutions lack mechanisms to bridge data and pedagogy, limiting their ability to support real-time, evidence-based interventions in the classroom \cite{sarmiento_participatory_2022}.  

To address these gaps, UCD in educational technology increasingly incorporates data storytelling principles to enhance comprehension, usability, and decision-making. Studies have demonstrated that dashboards enriched with interactive visualizations, on-demand contextual explanations, and AI-driven summaries can reduce cognitive overload and improve the interpretability of learning analytics \cite{echeverria_towards_2017}. For example, narrative elements such as descriptive titles, annotated data points, and system-generated insights help teachers focus on key student behaviors and performance trends rather than manually analyzing complex datasets \cite{echeverria_exploratory_2018}.

\subsection{Gaze Analytics in Education and Learning Contexts}

Gaze behavior offers a rich stream of data for understanding student attention, strategy use, and cognitive engagement, especially in domains like reading, writing, and problem solving \cite{taylor_eye_2016, liu_l2_nodate}. In particular, eye-tracking has been widely used in educational psychology to explore how learners allocate attention across visual stimuli such as diagrams, text passages, and interface elements \cite{wang_towards_2021}. Within the field of learning analytics, gaze has helped uncover reading comprehension strategies, engagement patterns, and even metacognitive skills \cite{meziere_scanpath_2024, hutt_automated_2019}.

In ELA instruction, gaze is particularly promising because student performance is often difficult to observe during silent reading \cite{Busjahn2015EyeOrder}. Teachers must infer attention, confusion, or disengagement from limited cues. Gaze analytics could augment these inferences by offering concrete visualizations of where students looked, how long they fixated, and how their attention shifted during a task. Prior work has demonstrated that such gaze features can be predictive of comprehension or learning gains \cite{RajendranPredictingBehavior}, but translating those patterns into teacher-friendly feedback remains a challenge.

Most gaze-based systems in education remain research prototypes, with limited focus on real-world implementation or classroom usability \cite{davalos_gazeviz_2024}. Moreover, the dashboards used to present gaze data are often static, overly technical, or lack pedagogical framing, making them inaccessible to non-experts \cite{jayawardena_advanced_2024}. This motivates the need for more interpretable, context-aware gaze analytics tools that can support classroom stakeholders in making informed decisions.

\subsection{Conversational Agents and the Interpretation of Gaze Analytics}

The rise of generative AI (GenAI), particularly large language models (LLMs), has catalyzed the development of conversational agents for education \cite{davar_ai_2025}. These systems increasingly assist with tasks such as automated feedback generation \cite{Cohn_Hutchins_Le_Biswas_2024}, real-time student support \cite{molina2024leveragingllmtutoringsystems}, and data-driven instructional guidance \cite{HWANG2020}. By enabling natural-language interaction, conversational agents reduce the barrier to accessing complex learning data, especially for non-technical users like classroom teachers.

Among emerging tools, systems like VizChat exemplify how conversational interfaces can enhance learning analytics dashboards by providing interactive, on-demand explanations of student performance data \cite{yan_vizchat_2024}. Rather than relying solely on static charts or prewritten summaries, these agents allow educators to query trends, explore outliers, and receive tailored insights through dialogue. This shift reflects a broader movement toward accessible data storytelling and just-in-time support within EdTech systems.

However, most educational conversational agents to date operate on structured log data, such as quiz scores, clickstreams, or engagement logs, and have yet to incorporate richer, multimodal signals like eye-tracking \cite{davalos_llms_2025}. While gaze data has demonstrated potential for revealing student attention, reading behaviors, and comprehension strategies \cite{meziere_scanpath_2024,hutt_automated_2019}, its abstract nature makes it difficult for users to interpret without scaffolding. This represents a critical gap in both AI-driven feedback and learning analytics design.

Integrating gaze into LLM-powered conversational agents introduces new challenges around interpretability, relevance, and trust. Teachers must be able to verify the agent’s explanations and understand how its insights map back to observable learning behaviors. Without transparency, LLM outputs risk generating hallucinated or pedagogically misaligned conclusions \cite{nguyen_black_2023, weitz_let_2021}. As such, explainable AI (XAI) principles, such as traceability, confidence indicators, and source citations, are essential for maintaining credibility and instructional value \cite{hostetter_xai_2023}.

Our work contributes to this emerging space by exploring how conversational agents can be designed to interpret gaze-based analytics in an intelligible, pedagogically grounded manner. We investigate how LLMs can not only summarize gaze features into human-readable reports but also support real-time, queryable feedback via interactive agents. This integration bridges a critical gap between novel multimodal learning data and everyday classroom decision-making.

\section{Research Approach}

\begin{figure}[h]
    \centering
    \includegraphics[width=\linewidth]{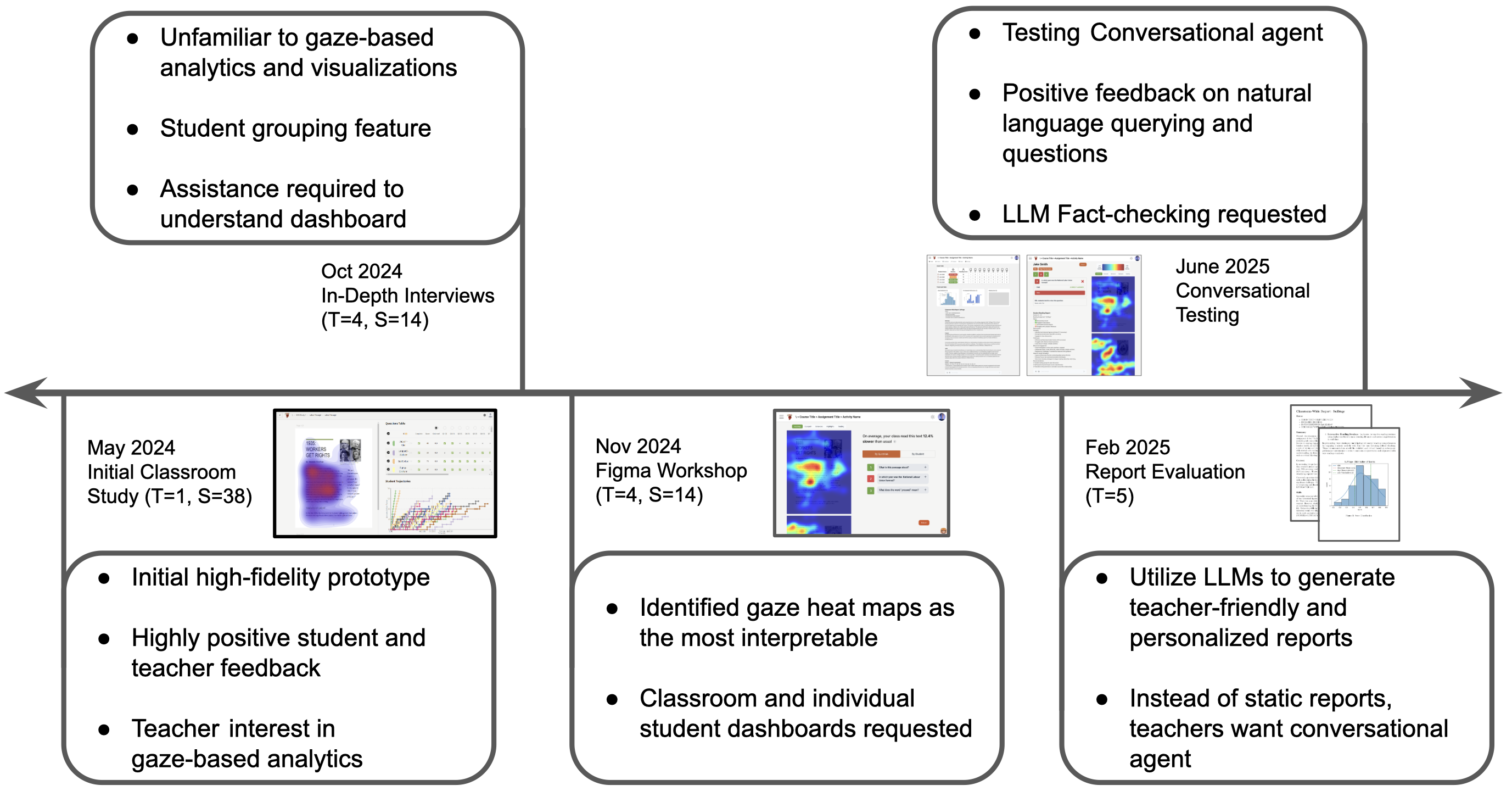}
    \Description{
    A horizontal timeline showing five iterative research phases from May 2024 to June 2025, each with a labeled milestone, participant counts, and key insights. 
    
    - In May 2024, the Initial Classroom Study (T=1, S=38) tested a high-fidelity prototype and received positive feedback from students and teachers. Teachers expressed interest in gaze-based analytics. 
    
    - In October 2024, In-Depth Interviews (T=4, S=14) revealed that participants were unfamiliar with gaze visualizations, requested a student grouping feature, and needed assistance interpreting the dashboard.
    
    - In November 2024, a Figma Workshop (T=4, S=14) identified heatmaps as the most interpretable visualization. Teachers requested both classroom and individual student dashboards.
    
    - In February 2025, during Report Evaluation (T=5), teachers evaluated LLM-generated reports. They appreciated the personalized insights but preferred a more interactive conversational agent over static summaries.
    
    - In June 2025, during Conversational Testing, teachers tested the integrated agent, gave positive feedback on natural language querying, and requested fact-checking features to verify LLM responses.
    
    Small screenshots of dashboard designs and gaze visualizations accompany each milestone.}
    \caption{\textbf{User-Centered Design Timeline}: Key phases of our iterative design process, including teacher and student involvement and major takeaways that shaped each prototype revision.}
    \label{fig:timeline}
\end{figure}

Our work follows a design-based research (DBR) methodology \cite{anderson_design-based_2012, the_design-based_research_collective_design-based_2003} centered on iterative refinement and stakeholder involvement throughout the development of a gaze-based learning analytics dashboard for ELA instruction. Rather than relying on a fixed hypothesis or a linear experimental design, this research embraced a cyclical, UCD process in which system functionality, usability, and pedagogical relevance were continuously evaluated and improved in collaboration with teachers and students.

This approach reflects the complexities of educational technology development, where the effectiveness of a system depends not only on its technical capabilities but also on its alignment with classroom practices, teacher workflows, and student engagement \cite{joseph_practice_2004}. By involving stakeholders at multiple stages, through classroom deployments, in-depth interviews, design workshops, and evaluations, we ensured that the dashboard evolved in response to real-world feedback rather than speculative assumptions. At each data collection, all included participants -- students and teachers alike -- debriefed on the study's object and their written consent (including the student's parents) were collected while following the guidelines of the \school IRB. Each iteration of the system was both informed by and tested against classroom needs, contributing to a holistic and practice-grounded development process.

As shown in Figure~\ref{fig:timeline}, our research progressed through four key stages:

\begin{enumerate}
    \item \textbf{Initial Classroom Study:} A feasibility study using a high-fidelity prototype tested whether gaze-based analytics could be effectively collected and interpreted in a live classroom setting. Feedback from this deployment shaped early understanding of how gaze visualizations might support student reflection and teacher decision-making.
    
    \item \textbf{In-Depth Interviews with Teachers and Students:} We conducted qualitative interviews to better understand how users interpreted the prototype, what challenges they faced, and how the design could better support instructional and learning needs. This phase generated critical insights that led to revised interface components and new features, such as visual aids and grouping functionalities.
    
    \item \textbf{Interactive Design Workshop:} Using an updated prototype populated with sample data, we held an in-person co-design session to evaluate new features, gather preferences for data representation, and validate emerging design choices. Participants offered actionable feedback on elements such as visualization types, interface layout, and information density.
    
    \item \textbf{Conversational Agent Integration:} In response to the recurring need for interpretability and workload reduction, we developed an LLM-driven reporting system and integrated a conversational agent into the dashboard. This conversational layer was then evaluated through targeted teacher testing sessions to assess its usefulness, accuracy, and potential for supporting on-demand inquiry and sense-making.
\end{enumerate}

At each stage, design decisions were informed by empirical findings from previous phases. This allowed for the continuous evolution of the dashboard’s functionality, usability, and pedagogical value. Our approach reflects a commitment to collaborative design and educational relevance, with the goal of creating a system that not only visualizes student gaze behavior but also enhances instructional decision-making and learning outcomes in the classroom.

\section{Iterative Design Process}

\begin{figure}[h]
    \centering
    \includegraphics[width=\linewidth]{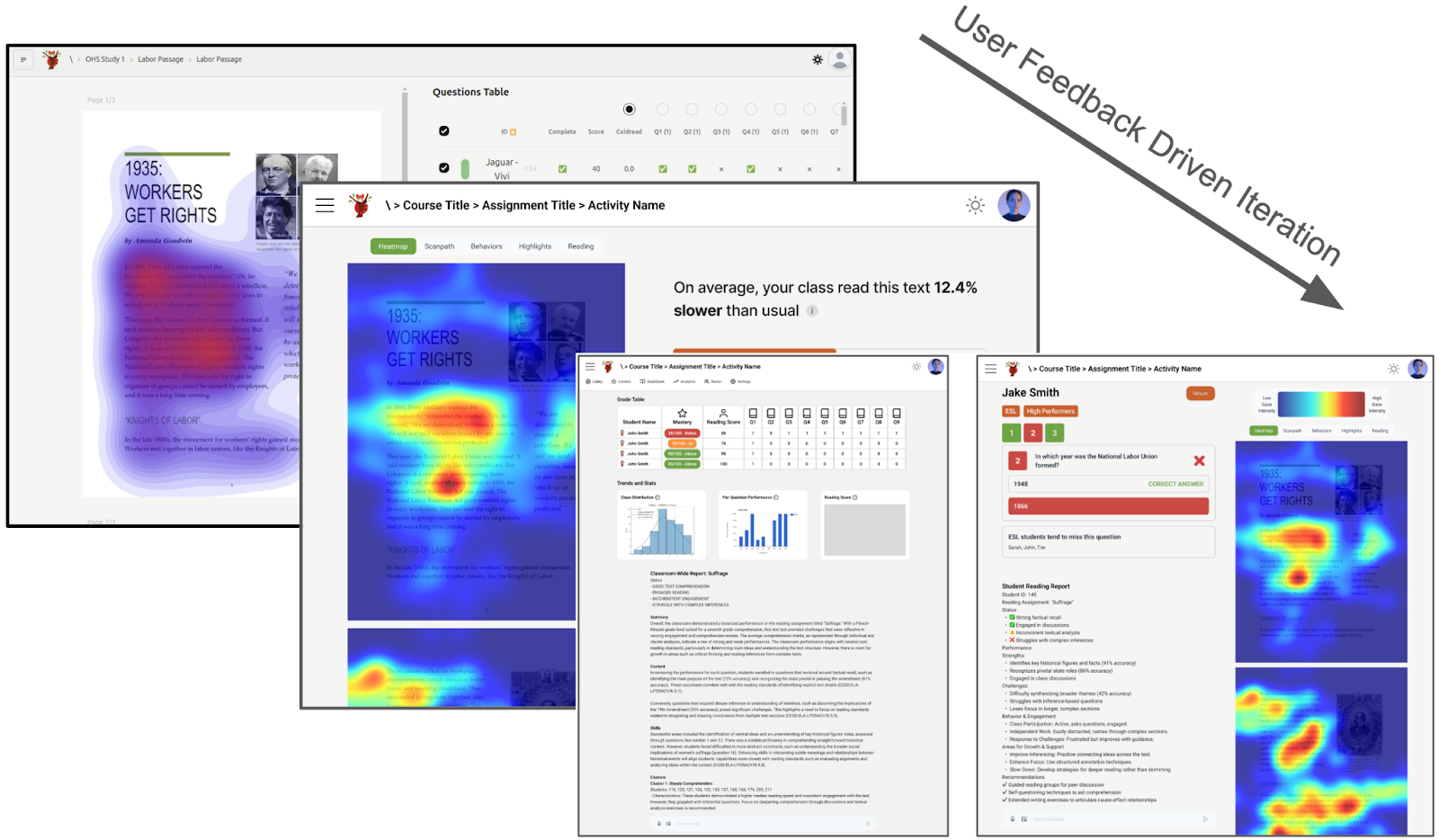}
    \Description{
    A sequence of dashboard mockups created in Figma, showing the visual progression from early static wireframes to interactive, high-fidelity prototypes. The designs reflect iterative improvements such as clearer layouts, gaze heatmaps, accordion-style question views, and the integration of conversational agents, based on teacher and student feedback.}
    \caption{\textbf{Figma-Guided Design Evolution}: Figma served as a central collaborative artifact throughout the project, supporting cross-team discussions, rapid prototyping, and stakeholder engagement. Each iteration of the dashboard incorporated user feedback, evolving from static wireframes to interactive prototypes informed by real classroom needs.}
    \label{fig:figma_design}
\end{figure}

A central component of our iterative design methodology was the use of Figma as a collaborative workspace to drive communication, reflection, and refinement throughout the development of the gaze-analytics dashboard. From the outset, we recreated the core interface of the \learningEnvironment system within Figma to establish a shared visual language and create a tangible design artifact for both internal design discussions and external stakeholder engagements. This artifact enabled the research team to ground conversations with teachers and students in a concrete interface, reducing ambiguity and fostering more targeted, context-aware feedback.

As the project evolved, our use of Figma transitioned from static mockups to fully interactive prototypes that allowed for the simulation of real system behaviors, such as collapsible accordions, hover states, and dynamic legends. This shift enabled richer exploration of design ideas, allowing us to quickly test and revise UI elements based on stakeholder input. Each iteration reflected a co-design philosophy, incorporating the voices of educators and students directly into visual and structural changes. The result was a living prototype that not only supported participatory design practices but also streamlined alignment across our interdisciplinary team of researchers, designers, and domain experts.


\subsection{Initial Classroom Study}

\begin{figure}
    \centering
    \includegraphics[width=\linewidth]{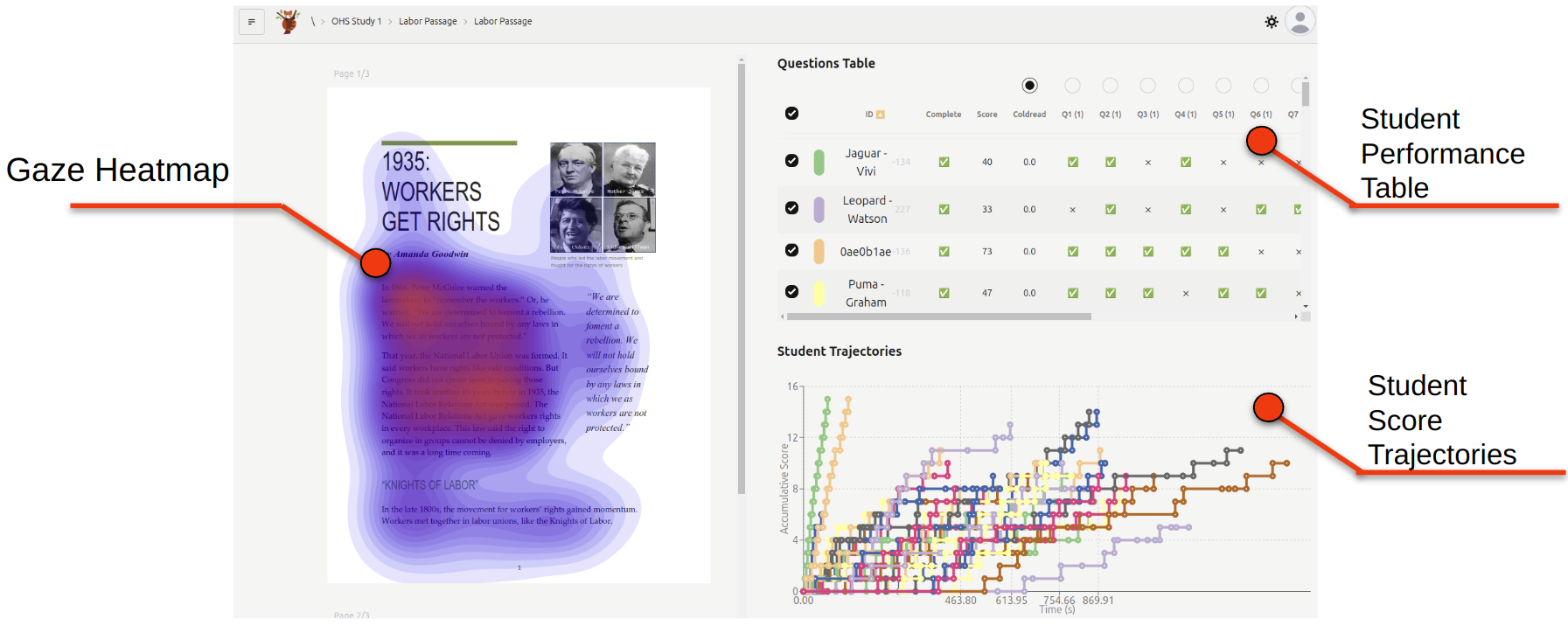}
    \caption{\textbf{Post-Assessment High-Fidelity Prototype v1}: The initial dashboard design used during the first classroom study. The interface presents three key components: a gaze heatmap overlaid on a reading passage (left), a performance table showing student answers per question (top right), and a score trajectory graph visualizing cumulative student scores over time (bottom right). This version was used to assess student engagement with gaze visualizations and inform early system refinements.}
    \Description{
    Screenshot of an early prototype dashboard. On the left, a reading passage is overlaid with a purple gaze heatmap showing areas of student visual attention. On the right, a table lists student names and their question-by-question scores, with checkmarks and Xs. Below the table, a colorful line graph displays student score trajectories over time, indicating how their cumulative scores changed during the assessment. Red annotations label the gaze heatmap, student performance table, and score trajectories.}
    \label{fig:version1}
\end{figure}

\paragraph{Objective} 
The initial phase focused on evaluating the feasibility of bring eye-tracking technology to the classroom and exploring whether gaze visualizations could meaningfully support student reflection and teacher decision-making, laying the groundwork for further design, iteration, and eventual integration into a fully-scalable feedback system powered by a conversational agent.

\paragraph{Protocol}
Using a high-fidelity prototype, we conducted a classroom study in May 2024 with one 5th-grade teacher (T=1) and 38 students (S=38; 22 male, 16 female) at a public elementary school in the southeastern United States. As part of the study, students completed a 500-word reading comprehension assessment within the \learningEnvironment system. Throughout the task, their gaze data was captured in real-time, stored securely, and processed to generate post-assessment analytics. Upon completion, both students and the teacher were given immediate access to a dashboard visualizing individual and class-level gaze-based metrics. Following the activity, students completed a short questionnaire designed to capture their perceptions of usability, clarity, and engagement with the system.

\input{figures/initial_classroom_study}

Additionally, at the end of the study, via an informal interview, the teacher was presented with the dashboard along with all students' data and asked questions regarding the interpretability, ease-of-use, and classroom utility. 

\paragraph{Insights}
The deployment was successful, with stable online gaze data collection and immediate presentation of analytics through version v1 of the post-assessment dashboard. In post-session questionnaires, most students (62\%, n=23) reported no significant challenges using the system. The remaining students (38\%, n=14) noted minor difficulties, including trouble navigating back to previously read content (n=2), locating specific interface elements (n=2), eye strain (n=1), and page formatting issues (n=1). These reports suggest that while the system was generally accessible, minor usability issues, particularly around layout and navigation, should be addressed in future iterations.

The majority of students (90\%) expressed positive sentiments toward the system, with many highlighting the novelty of the eye-tracking feature. Specifically, 76\% (n=28) mentioned the gaze-tracking component as particularly engaging, noting it was ``cool'' or ``fascinating'' to see how the system visualized their reading behaviors. Several students also appreciated unique elements of the interface, such as color-coded visualizations and the ability to revisit sections of the passage. These findings indicate strong student engagement with the concept of gaze-based feedback, validating the potential of eye-tracking as a reflective learning tool. However, the minor usability concerns highlight the need for improved interface clarity and accessibility, particularly for younger users.

\begin{figure}
    \centering
    \includegraphics[width=\linewidth]{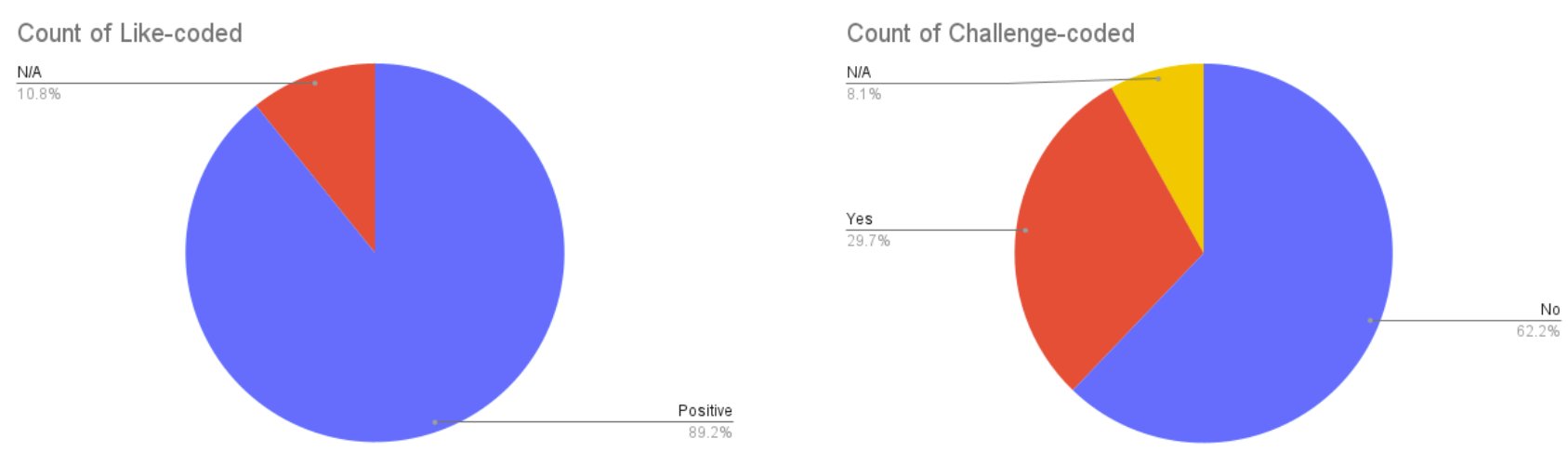}
    \caption{\textbf{Student Questionnaire Responses}: Pie charts summarizing student feedback from the initial classroom study. The left chart shows that 89.2\% of students gave positive comments about the system, while the right chart indicates that 29.7\% of students reported experiencing some form of challenge. These insights helped guide early usability refinements. Reproduced from \cite{davalos_gazeviz_2024} with permission.}
    \Description{
    Two pie charts showing student feedback responses. The first chart on the left, titled “Count of Like-coded,” shows 89.2\% positive responses, 10.8\% marked N/A. The second chart on the right, titled “Count of Challenge-coded,” shows 62.2\% no challenges, 29.7\% reported challenges, and 8.1\% marked N/A. The charts are color-coded: blue for majority segments, red for challenges or N/A, and yellow for other responses.}
    \label{fig:initial_study_questionnaire}
\end{figure}

In the exit interview, the teacher noted that the dashboard included promising analytics that could support both learning outcomes and the underlying learning processes. The teacher emphasized that such insights could be used to better praise students, tailor support, and adapt instructional practices throughout the course. However, concerns were raised about the dashboard’s complexity and the overwhelming volume of data. Additionally, the teacher acknowledged that most educators are unfamiliar with gaze-based analytics and recommended that future versions of the system better support data visualization literacy to convert complex visualizations into actionable, pedagogically meaningful feedback.

Our first classroom study supports a positive answer to \textbf{RQ1}: both students and the teacher found gaze-based analytics valuable and reflective of actual reading behaviors. Students engaged with the visualizations and used them to better understand their successes and mistakes. The teacher noted the potential for supporting adaptive instruction and student feedback, though also highlighted the need for improved data visualization support. Overall, the study suggests that gaze analytics, when well-designed, can be both approachable and actionable for classroom use.

\subsection{Interviews with Teachers and Students}

\input{figures/in_depth_interview_student_questions}
\input{figures/in_depth_interview_teachers}

\paragraph{Objective}

Building on insights from the initial classroom study, which demonstrated the feasibility and pedagogical promise of gaze-based analytics, the goal of this phase was to gather more detailed feedback from teachers and students on how to improve the system’s usability, interpretability, and instructional relevance. Specifically, we sought to understand how key gaze-based insights could be more effectively communicated through the dashboard to support both formative assessment and ELA pedagogy.

Grounded in our existing prototype, these interviews aimed to identify what types of information users would find most useful during their teaching or learning journeys, how they preferred that information to be visualized, and what role gaze data could play in shaping classroom instruction. In addition to interface-level feedback, we also explored broader educational needs, including the relationship between assessment design and learning analytics interpretation, to uncover opportunities for deeper integration and synergy across the instructional pipeline.

\paragraph{Protocol}
In early November 2024, we conducted separate in-depth interviews via Zoom with teachers (T=4) and students (S=14) for 30 minutes each. Audio recordings, transcripts, and observational notes were collected in accordance with participants' consent and \school IRB-approved privacy protocols. Interview questions were tailored to the ELA context and grounded in the current version of the \learningEnvironment system to ensure relevance and technical clarity. To avoid overly abstract or out-of-scope responses, we centered our discussions around the high-fidelity prototype, anticipating that teachers and students may struggle to propose enhancements to a modality as novel as gaze analytics. The interviews were guided by a presentation that displayed our questions alongside screenshots of the prototype, providing a visual artifact to stimulate conversation and elicit concrete feedback. Participants were asked to reflect on the system’s current features and suggest potential improvements to better support their teaching or learning. The full list of interview questions are shown in Tables \ref{tab:interview_students} and \ref{tab:interview_teachers} along with the interviews' presentation materials has been made publicly available at \href{https://drive.google.com/drive/folders/1YQAxujXG6WF9TYGQfxfh2fDEuRRnF4LP?usp=sharing}{https://drive.google.com/drive/folders/1YQAxujXG6WF9TYGQfxfh2fDEuRRnF4LP?usp=sharing}.

\paragraph{Insights}
Through our initial interviews with teachers and students, guided by the high-fidelity prototype, we identified a number of potential improvements to the post-assessment dashboard. One major finding was the perceived usefulness of an AI chatbot, especially in helping users interpret dashboard data. Participants expressed a desire for visual aids, legends, figure descriptions, and AI-generated summaries to improve the clarity of visualizations. Teachers specifically requested a student grouping feature to reduce the effort required to analyze and navigate classroom-level data. These groups, they suggested, should be flexible and teacher-defined, enabling instructors to organize students based on relevant educational categories, such as English as a Second Language (ESL) learners or performance tiers like Below, At, and Above Mastery. These insights guided modifications to the Figma design to enhance the dashboard’s usability and responsiveness to user needs.

\subsection{Interactive Workshop}

\begin{figure}
    \centering
    \includegraphics[width=\linewidth]{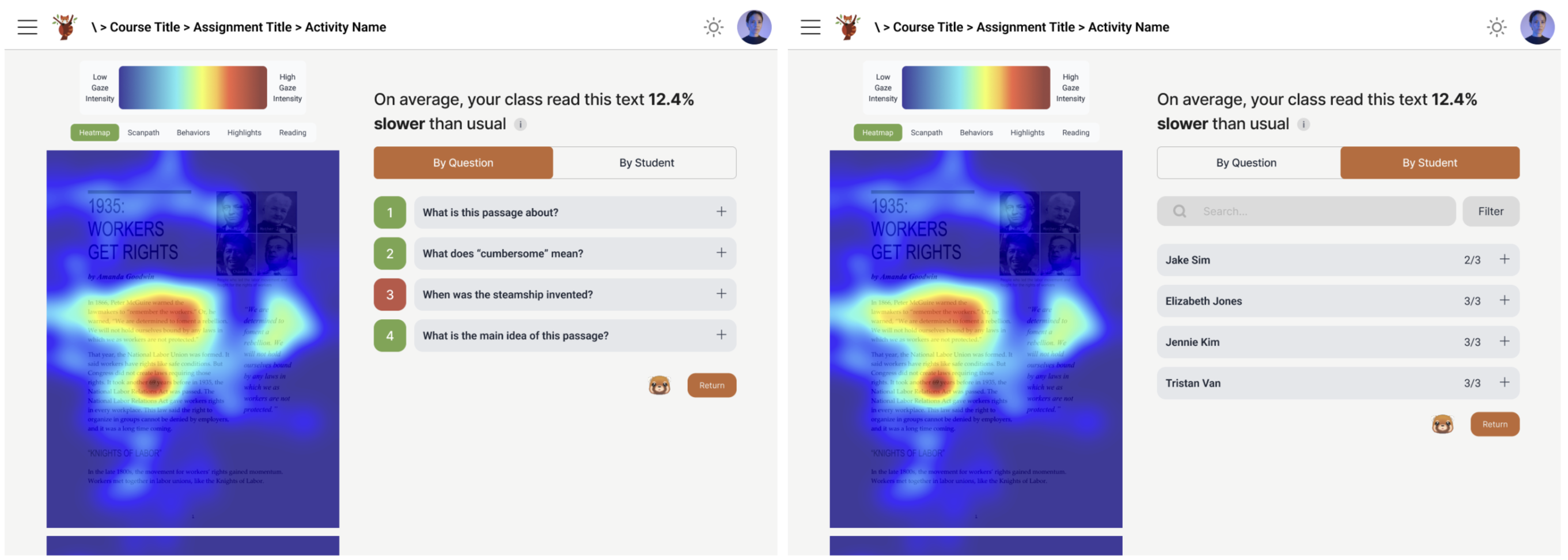}
    \caption{\textbf{Post-Assessment Figma Design v2}: Revised dashboard prototype shown during the interactive design workshop. This version includes an accordion-based question layout, class performance summaries, and PDF-embedded heatmaps. Participants could toggle between viewing data by student or question, with contextual prompts and updated interface structure.}
    \Description{
    Two side-by-side screenshots of a dashboard prototype. Both show a reading passage with a colorful gaze heatmap overlaid on it. The left screenshot features expandable comprehension questions; the right screenshot shows a list of students with reading scores. Both have navigation tabs for toggling between heatmaps, scanpaths, behavior views, and highlights.}
    \label{fig:interactive_figma_design}
\end{figure}

\begin{figure}
    \centering
    \includegraphics[width=\linewidth]{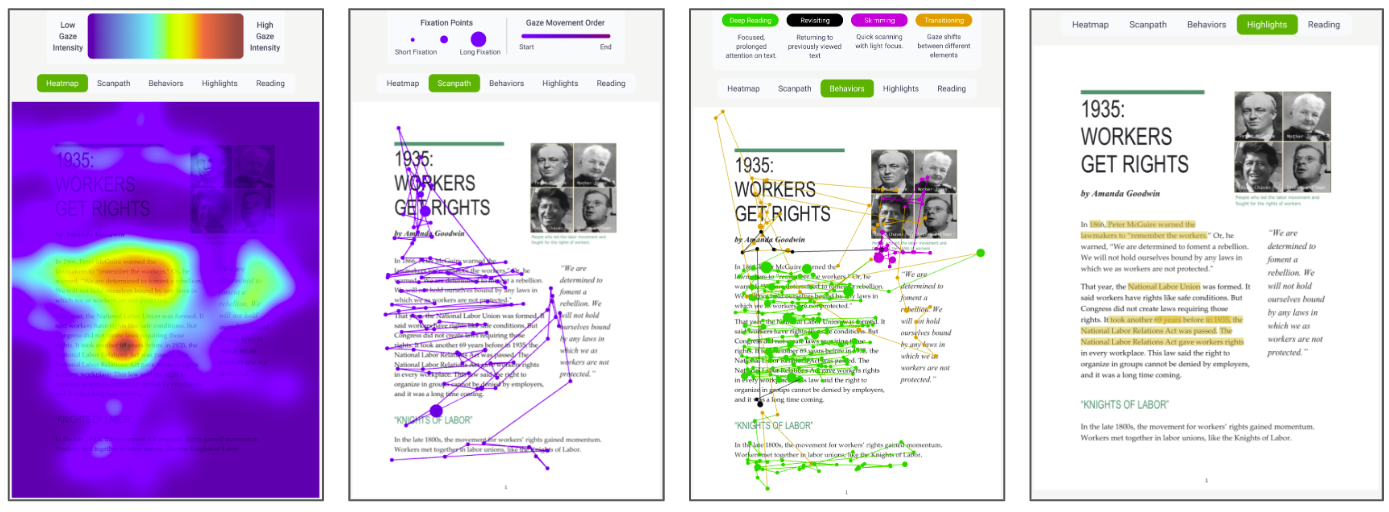}
    \caption{\textbf{PDF Overlay Visualization Types}: Comparison of four gaze-based visualization styles explored during design iterations. From left to right: (1) gaze heatmap shows intensity of attention, (2) raw scanpath visualizes fixation order, (3) behavior-segmented scanpath uses color-coded reading behaviors, and (4) highlight mode emphasizes key content areas.}
    \Description{
    Four side-by-side visualizations of the same reading passage. From left to right: a colorful heatmap showing gaze intensity; a scanpath of numbered dots with arrows indicating fixation order; a color-coded behavior scanpath distinguishing reading strategies; and a version of the passage with highlighted sentences. Each has navigation tabs for switching between modes.}
    \label{fig:pdf_visualization}
\end{figure}

\paragraph{Objective}  
Following the interview phase, we aimed to validate and refine our design updates through an interactive workshop that brought teachers and students into direct engagement with a revised, highly-interactive Figma prototype shown in Fig. \ref{fig:interactive_figma_design}. This iteration incorporated several participant-driven improvements, including clearer legends, accordion-style layouts for question-level analysis, and flexible student grouping and tagging functionalities tailored for classroom-level data exploration. A key objective of this workshop was to assess the usability and interpretability of new gaze visualizations overlaid on PDF reading materials, including heatmaps, raw scanpaths, and behavior-segmented scanpaths. We also sought feedback on the placement and utility of the newly integrated AI chatbot, as well as the overall clarity and navigability of the updated dashboard design. This phase allowed us to evaluate how well the revised interface met user needs and to gather further insights for guiding subsequent iterations.

\paragraph{Protocol}
We conducted an in-person, interactive design workshop involving 20 students and 4 teachers, each interacting with a high-fidelity Figma prototype populated with representative data. Participants were split into small groups, 1 researcher per 5 students or 2 teachers, to allow for active scaffolding and observation. Each participant accessed the interactive prototype independently, while researchers guided them through a structured set of tasks and discussion prompts.

The workshop followed a three-stage protocol framework, \textit{Observe} $\rightarrow$ \textit{Action} $\rightarrow$ \textit{Feedback}, to scaffold participant engagement with the interface. During the \textit{Observe} phase, participants were asked to explore and describe interface components to build familiarity. The \textit{Action} phase prompted them to complete targeted tasks to assess the dashboard’s usability and interpretability. Finally, the \textit{Feedback} phase invited reflections on usability, clarity, and desired features. Sessions were organized into four core areas of interaction: Course Overview, Assignment Editing (teachers), Assignment Activity (students), and Post-Assessment Dashboard. For each, participants engaged with specific components such as visualizations, chatbot integration, annotation tools, and gaze-based feedback. Researchers collected observation notes, clarifying questions, and emergent suggestions throughout the sessions. The full protocol is detailed in Table~\ref{tab:figma_workshop_questions}.

\input{figures/figma_workshop}

\paragraph{Insights}

The interactive workshop surfaced a range of insights that helped validate, critique, and refine our updated dashboard design. Both students and teachers responded positively to several interface improvements, particularly the accordion-style layout for question-level feedback, clearer legends, and the overall reduction of visual clutter. These updates contributed to smoother navigation and more digestible information flow. However, feedback also revealed a number of areas that required further refinement, especially in terms of visualization interpretability, information hierarchy, and workload distribution.

A central takeaway from this workshop was the need to incorporate \textbf{data storytelling principles} to improve how gaze-based analytics are communicated to students and teachers. Participants consistently favored visualizations and features that framed data within a meaningful and interpretable narrative. For example, heatmaps emerged as the most intuitive and well-received gaze visualization across both groups. Students described them as ``easy to understand'' and ``familiar,'' particularly when overlaid on the original reading text, in contrast to scanpaths and behavior-segmented plots, which were seen as confusing or overly complex. These preferences suggest that storytelling in gaze analytics begins with selecting visual forms that align with user expectations and cognitive comfort—principles echoed in narrative visualization research.

To support better interpretability, participants proposed concrete storytelling aids such as simplified legends, info buttons, behavioral summaries, and dynamic tooltips to contextualize the meaning of visual elements. Teachers stressed that for advanced visualizations to be pedagogically useful, they must include domain framing, such as explanations of how behaviors relate to comprehension strategies or learning standards. These findings affirm that gaze analytics in education benefit from \textbf{scaffolded interpretation}, where design elements explicitly guide the user toward understanding rather than requiring them to decode unfamiliar data representations.

A key theme that further aligned with data storytelling principles was the \textbf{need for personalization and self-referenced feedback}. Students overwhelmingly preferred dashboards that highlighted their individual progress over time rather than comparative class-wide data. They requested longitudinal trends, improvement metrics, and specific guidance (e.g., ``What can I do to improve?'')—all components that help construct a coherent narrative around their own learning journey. Teachers similarly valued features that allowed flexible grouping and filtering to generate \textbf{instructionally relevant “stories”} about student needs, such as grouping by ELL status or mastery levels. Together, these insights underscore the importance of designing dashboards not only to present data but to help users construct meaningful narratives from it.

Additional suggestions emphasized usability enhancements that further support storytelling through clarity and relevance. These included: (1) larger fonts and improved layout for readability, (2) zoom features for detailed inspection, (3) toggles to control annotation visibility, and (4) summaries of key reading metrics like skipped questions or time-on-task. Both teachers and students expressed strong interest in \textbf{question-level narratives}, such as linking a comprehension question directly to its referenced passage section or categorizing questions by cognitive skill (e.g., inferential vs. literal). These design requests reinforce the need for content-tagging and task-type analytics that help narrativize performance through domain-relevant lenses.

Feedback on the newly integrated conversational agent was similarly tied to data storytelling goals. Students appreciated having a contextualized and collapsible chatbot to support interpretation, while teachers valued the agent’s ability to explain complex dashboard elements. However, teachers emphasized the need for transparency and traceability, requesting that the agent explicitly cite data sources or point to underlying patterns that support its summaries, echoing best practices in explainable AI and narrative accountability.

Finally, both groups stressed that the dashboard still placed too much cognitive load on users to manually interpret insights. Teachers requested a \textbf{top-down narrative structure}: starting with a concise classroom overview and then drilling into detailed student-level or question-level narratives as needed. A text-based classroom report was suggested as an approach to complement the data visualizations and establish a clear and actionable narrative. Students expressed interest in on-boarding tutorials or guided walk-throughs that could help them understand how to navigate and interpret the dashboard, suggesting the utility of \textbf{narrative scaffolds} even at the level of interface interaction.

Overall, the workshop validated several design decisions from the interview phase while highlighting critical opportunities to better align with the principles of data storytelling. These include prioritizing intuitive visual forms, supporting self-narration through personalization, embedding explanatory scaffolds, and guiding users through structured analytic journeys. These insights directly shaped the next iteration of the system, including refined gaze visualizations, enhanced legends and info buttons, contextualized chatbot support, and a layered dashboard architecture designed to reduce cognitive overload while increasing pedagogical relevance, advancing our ability to answer \textbf{RQ2}.

\subsection{LLM-Generated Reports and Conversational Agent}

\begin{figure}
    \centering
    \includegraphics[width=\linewidth]{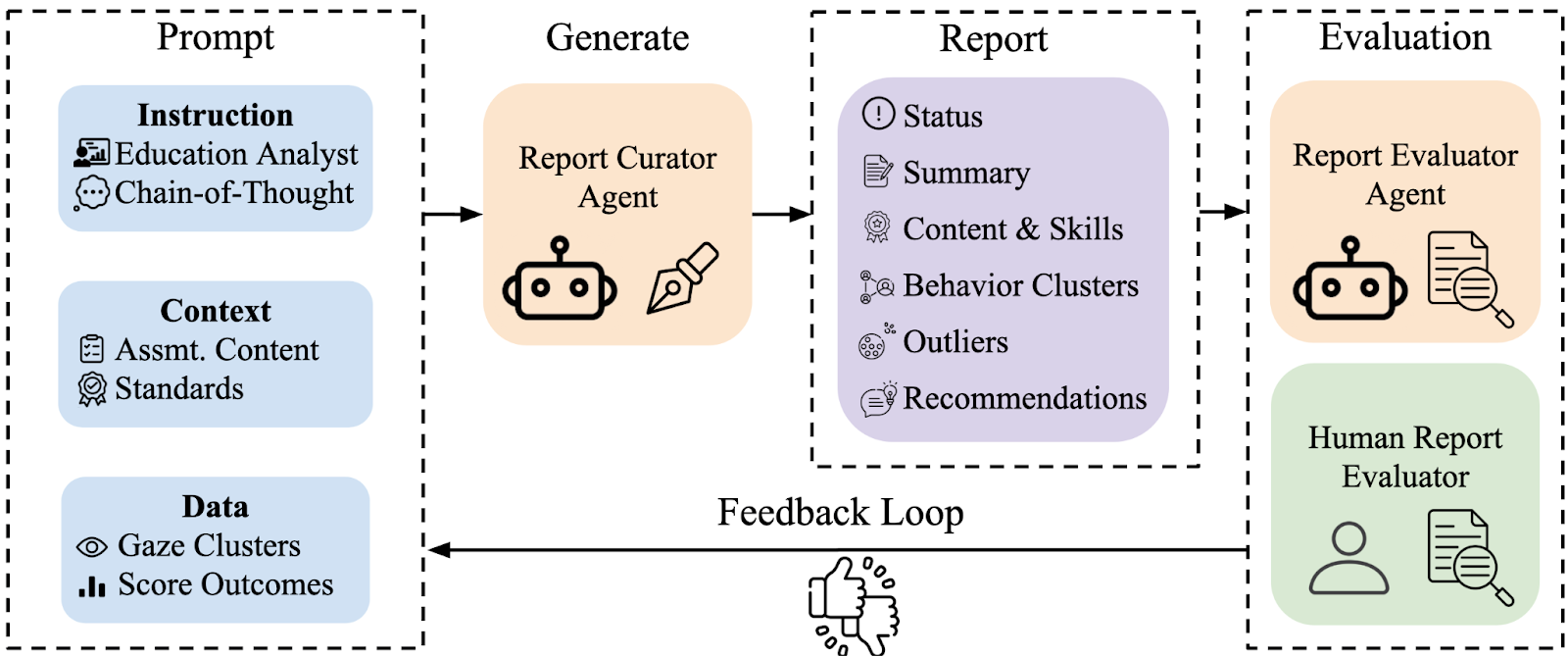}
    \caption{\textbf{LLM-Augmented Report Generation Pipeline}: Gaze data, student performance, assignment content, and learner attributes are aggregated and structured into a prompt format used by a large language model (LLM) to generate classroom-level summaries. These human-readable reports highlight key behavioral patterns and instructional insights, serving as the foundation for both static feedback and interactive, conversational explanations within the dashboard. Reproduced from \cite{davalos_llms_2025} with permission.}
    \Description{
    Diagram showing four main stages of an LLM-powered feedback system. On the left, a "Prompt" block includes instructions, assignment context, and input data like gaze clusters and scores. This feeds into a "Generate" block with a Report Curator Agent that produces a structured "Report" containing status, summaries, content skills, behavior clusters, outliers, and recommendations. The report is then evaluated by an LLM-based Report Evaluator Agent and a Human Evaluator. A feedback loop sends insights from evaluation back to improve future prompt construction.}
    \label{fig:pipeline}
\end{figure}

\paragraph{Objective}
In response to teacher requests for more accessible and actionable insights, we developed an LLM-driven assignment report system that generates personalized classroom summaries using gaze-based behavioral clusters and relevant assignment components. These components include the assignment text, question sets, and aligned ELA state standards, all of which serve as inputs for the language model to produce deeper, contextually grounded analyses. The system is designed to identify relationships between behavioral patterns and learning outcomes, offering teachers an overview of classroom performance along with descriptive profiles and explanations for each identified cluster of reading behaviors.

\paragraph{Protocol}
In February 2025, five teachers (T=5) reviewed a sample LLM-generated assessment report derived from gaze and performance data collected in prior study. The report was based on a reading comprehension assignment adapted from the National Assessment of Educational Progress (NAEP) and included sections on classroom status, learning goals, student clusters, outliers, and recommendations. Teachers evaluated each report section through a Likert-scale survey and provided open-ended feedback.

Following this initial evaluation, we developed a conversational agent prototype to explore how the LLM-generated report could be made more dynamic and responsive. In a follow-up study in June 2025, two additional teachers (T=2) were invited to explore the full analytics dashboard, now featuring the conversational agent, and ask questions about specific plots, student trends, or data interpretations. Feedback from this interaction focused on usability, interpretability, and trust in the LLM's responses.

\paragraph{Insights}

\begin{figure}
    \centering
    \includegraphics[width=\linewidth]{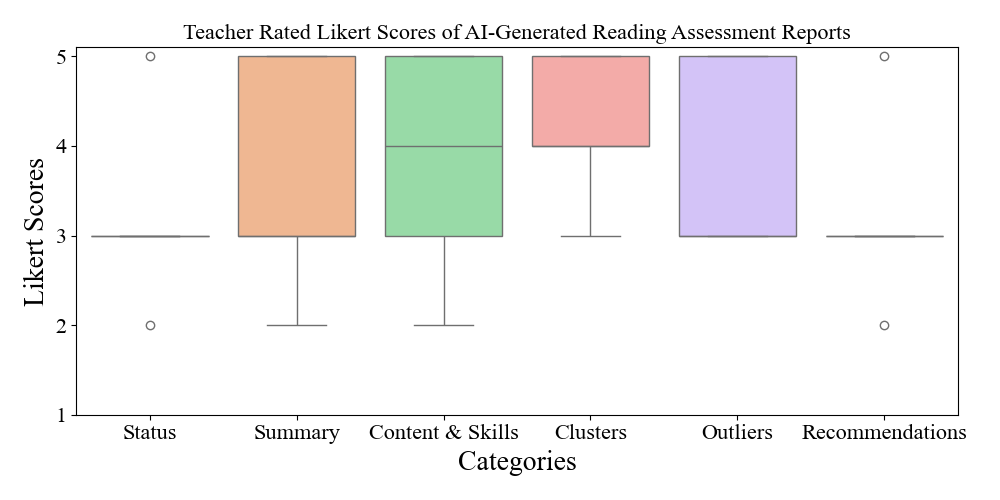}
    \caption{\textbf{Teacher Ratings of AI-Generated Report Sections}: Boxplot showing Likert scale ratings (1–5) provided by teachers for six different sections of the AI-generated reading assessment reports: Status, Summary, Content \& Skills, Clusters, Outliers, and Recommendations. While most sections received high median ratings, variability across responses highlights preferences for clarity, conciseness, and pedagogical utility.}
    \Description{
    A horizontal boxplot chart titled "Teacher Rated Likert Scores of AI-Generated Reading Assessment Reports." The x-axis lists six categories: Status, Summary, Content & Skills, Clusters, Outliers, and Recommendations. The y-axis ranges from 1 to 5, representing Likert scale ratings. Each category has a colored boxplot showing score distribution, with most medians between 3 and 5. Outliers are visible for several categories, especially Status and Recommendations.}
    \label{fig:teacher_feedback}
\end{figure}

Teachers responded positively to the structure and pedagogical relevance of the LLM-generated reports. They highlighted the usefulness of behavioral clusters for distinguishing different reading strategies and appreciated the alignment of insights with ELA standards. Teachers noted that the reports helped identify strengths and weaknesses across the classroom and could support instructional planning. However, several also commented that the reports were occasionally too verbose and requested more concise, skimmable summaries.

Most notably, teachers expressed strong interest in interacting with the report content rather than consuming it passively. They wanted to probe deeper into findings, generate follow-up queries, and ask the system for clarification—motivating the design of the conversational agent.

During the conversational prototype evaluation, both teachers found the agent helpful for interpreting visualizations and explaining unfamiliar gaze metrics. They particularly appreciated the ability to ask questions in natural language. However, concerns were raised regarding the factual reliability of the responses, especially when LLM-generated insights could not be directly verified against raw data. Teachers requested a mechanism to validate agent outputs and suggested the possibility of using the agent to conduct on-demand analysis in natural language, such as defining custom student groups or generating custom visualizations based on user-defined criteria.

\begin{figure}
    \centering
    \includegraphics[width=\linewidth]{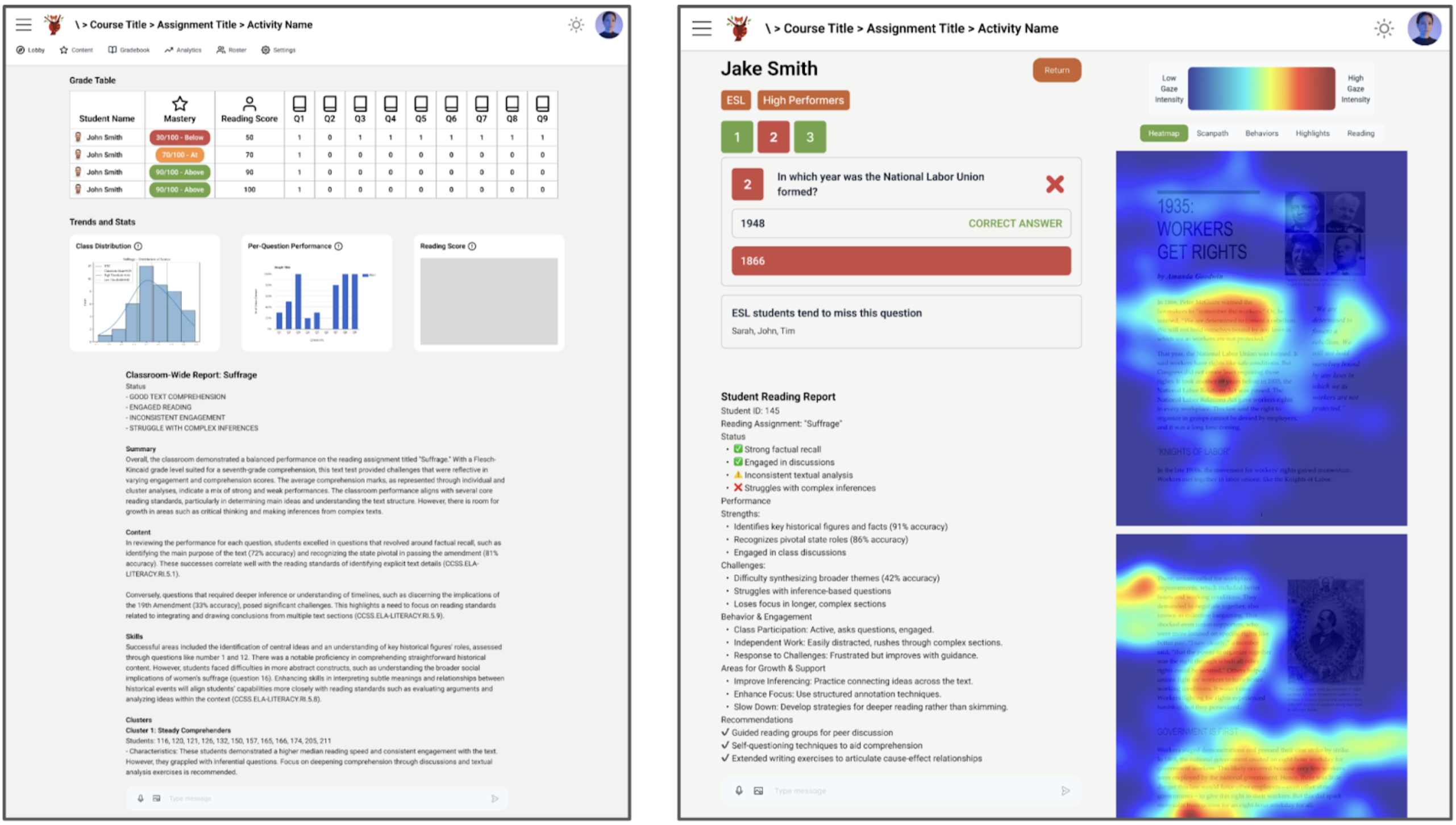}
    \caption{\textbf{Post-Assessment Analytics Dashboard Design v3}: The new version of the analytics dashboard includes both a classroom-wide (left) and an individual student (right) report. Each report has multimodal learning analytics and an AI-based report and conversational agent to support students and teachers use of the dashboards.}
    \Description{
    Two dashboard screenshots. The left side shows a classroom view with a grade table, bar charts for trends and stats, and a detailed classroom-wide report summarizing reading performance and behavioral clusters. The right side shows an individual student view with a marked reading question, personalized feedback, and two heatmaps overlaid on reading passages indicating gaze intensity. Text-based summaries include strengths, challenges, and recommended instructional strategies.}
    \label{fig:design_v3}
\end{figure}

\section{Findings and Implications}

In this section, we synthesize findings from five iterative design studies to answer our research questions and identify design implications for future gaze-based analytics systems. Our results point to both the promise and challenges of using gaze data in classroom-facing dashboards, especially when combined with data storytelling techniques and conversational agents.

\subsection{Summary of Findings}

\textbf{RQ1: Would teachers and students find gaze-based analytics approachable and capable of offering actionable insights?}  
Across all studies, both students and teachers expressed enthusiasm for gaze-based analytics. In the initial classroom deployment, students were curious and engaged, describing the gaze features as “cool” and “helpful,” while teachers saw potential in using gaze data to reveal hidden reading behaviors. However, both groups required additional support to interpret the visualizations effectively. While heatmaps were easily understood, scanpaths and behavioral visualizations proved too complex without supplemental guidance. These findings highlight the importance of not just collecting rich behavioral data, but also packaging it in ways that are cognitively approachable and pedagogically relevant.

\textbf{RQ2: How can we apply data storytelling principles to the design of gaze analytics tools for ELA instruction?}  
The application of data storytelling principles, such as clear narrative structure, contextual explanations, personalization, and guided interpretation, significantly improved the usability and pedagogical utility of the dashboard. Features like per-question summaries, student tagging, simplified legends, and interactive tooltips helped users make sense of complex data. The importance of personalization emerged strongly: students preferred to track their own progress over time rather than comparing themselves to peers, and teachers wanted dashboards to surface stories about specific instructional categories, such as English Language Learners or students below mastery. These findings demonstrate that storytelling in educational dashboards must be both data-driven and user-centered.

\textbf{RQ3: Could the use of a conversational agent assist in the navigation and interpretation of gaze data within an analytics dashboard?}  
The integration of a conversational agent into the dashboard enabled users to engage with the data more fluidly. Teachers appreciated the ability to ask natural-language questions and receive summaries or clarifications on gaze-based insights. However, issues of trust and transparency surfaced as central concerns. Teachers requested that AI-generated insights be grounded in verifiable data, and even suggested that agents should perform custom analyses on request (e.g., “Group students who scored above 80\% but read quickly”). These results suggest that conversational agents can reduce interpretive burden but must be designed with mechanisms to foster user trust, such as traceability, fact-checking, and explanation scaffolds.

\subsection{Design Implications}

Based on our multi-phase findings, we propose the following design principles for the development of gaze-based educational dashboards:

\begin{itemize}
    \item \textbf{Start with Familiarity}: Heatmaps were consistently preferred over scanpaths or abstract behavioral visualizations. Systems should lead with familiar, low-barrier visuals and reserve more complex representations for users with deeper interest or expertise.
    
    \item \textbf{Progressive Disclosure}: Dashboards should adopt a top-down structure, offering high-level summaries before revealing more granular insights. For example, a classroom-wide summary can guide teachers toward interesting outliers or trends, with drill-downs available for student-level or question-level analysis.

    \item \textbf{Support Self-Narration}: Students and teachers benefit from dashboards that highlight personal progress, growth over time, and tailored recommendations. Personalized narratives, such as “how I improved” or “where my students struggle”, can foster engagement and more meaningful reflection.

    \item \textbf{Layer in Storytelling Aids}: Features such as annotated visualizations, simplified legends, hover-based tooltips, and AI-generated summaries help translate gaze data into actionable narratives. These tools scaffold understanding without requiring prior expertise in eye-tracking.

    \item \textbf{Design for Explainable AI}: Conversational agents must be transparent and trustworthy. LLM-generated outputs should be traceable to underlying data, with confidence indicators or references to visual elements. Teachers should be able to challenge or verify system-generated claims.

    \item \textbf{Enable On-Demand Inquiry}: Agents should allow users to describe the analyses they want performed (e.g., “Compare ESL students’ heatmaps to non-ESL students”) and generate appropriate visualizations or summaries in response. This interactivity can bridge the gap between static analytics and exploratory pedagogy.
\end{itemize}

\subsection{Implications for Research}

Our work contributes to the growing field of explainable learning analytics by demonstrating how gaze data can be effectively surfaced to end-users in educational settings. The combination of UCD, narrative visualization, and conversational interaction provides a powerful framework for making novel data modalities accessible to non-experts. These findings extend previous research on learning analytics dashboards by showing how MMLA tools, when guided by storytelling principles, can become both interpretable and pedagogically valuable \cite{echeverria_driving_2018, cohn_multimodal_2025}.

In addition, this work highlights underexplored intersections between eye-tracking, generative AI, and human-computer interaction. While gaze and LLMs have individually been applied in educational contexts, their integration within a cohesive dashboard system represents a new frontier \cite{davalos_llms_2025}. Future research can build on this by studying how agents can not only interpret, but also generate new data-driven instructional strategies based on user queries and system insights.

\subsection{Limitations}

Our study has several limitations. First, although our design studies involved both teachers and students, the sample sizes, particularly for interviews and agent testing, were relatively small and may not generalize across all classroom contexts. Second, while webcam-based eye tracking has improved accessibility, it remains less precise than laboratory-grade equipment, which may limit the granularity of gaze features used in analytics. Third, while our conversational agent was well received, its performance depends on the reliability and interpretability of LLM outputs, which can be inconsistent and opaque.

Finally, our evaluations primarily focused on perceived usability and interpretability, rather than direct learning outcomes or long-term behavior change. Future work should explore how gaze-informed feedback affects instructional planning and student learning over time, including longitudinal studies that measure changes in pedagogical practice or comprehension outcomes.

\section{Conclusion}

This work presents the iterative design of a gaze-based learning analytics dashboard for ELA, developed through a series of classroom studies, interviews, design workshops, and AI-integrated prototypes. Our findings show that gaze data, while unfamiliar to most educators, can be made actionable and interpretable through a combination of user-centered visual design, data storytelling principles, and scaffolded interaction.

We further demonstrate the value of integrating conversational agents powered by large language models to support user understanding of multimodal analytics. These agents help bridge the gap between complex data and instructional decision-making, though concerns around trust and explainability remain. Together, our contributions highlight design strategies for future EdTech tools that seek to integrate novel data modalities like eye tracking in ways that are transparent, usable, and pedagogically relevant.

\section{GenAI Usage Disclosure}
GitHub Copilot was utilized to co-write the source code for this project. All generated code was thoroughly evaluated by human researchers, either corrected or removed to fulfill the purpose of our research analysis.

\bibliographystyle{ACM-Reference-Format}
\bibliography{mendeley_CHI2025,zotero_VU_Multimodal_Survey,manual,zotero_redforest,zotero_webeyetrack}

\end{document}

%% file: figures/initial_classroom_study.tex
\begin{table}[h]
    \centering
    \caption{Protocol for Initial Classroom Study}
    \begin{tabularx}{\textwidth}{l}
        \toprule
        Presentation and Introduction: 5 min\\
        Reading Comprehension Assessment: 15 min\\
        Student Questionnaire (questions below): 5 min\\
        Teacher Exit Interview: 15 min\\
        \hline
        \textbf{Student Questionnaire} \\
        1. Did you notice any challenges with the experience or the user interface? \\
        \hspace{1cm} a. What was confusing? \\
        2. What did you like about the interface? \\
        \bottomrule
    \end{tabularx}
    \label{tab:my_label}
\end{table}

%% file: figures/in_depth_interview_student_questions.tex
\begin{table}[]
\centering
\caption{\textbf{Student Interview Protocol and Questions}}
\begin{tabular}{p{2cm}|p{7cm}|p{5.5cm}}
\toprule
\textbf{Topics}                           & \textbf{Main Questions}                                                                                                                                                                                                                                                                                                                                                                                                                                                                 & \textbf{Follow-up questions / Prompts / Keywords to look for}                                                                                                                                                                                                                                                                                                              \\
\toprule
\multicolumn{3}{p{\textwidth}}{To start our conversation, we would like to first ask you questions regarding your experience with reading assignments}                                                                                                                                                                                                                                                                                                                                                                                                                                                                                                                                                                                                                                                                                                                                                                       \\
\hline
Current Experience                        & 1. What types of reading assignments and feedback do you receive?                                                                                                                                                                                                                                                                                                                                                                                                                       & 1a. How do you process this feedback? \newline 1b. Is this feedback important or helpful to you? \newline 1c. What kind of feedback did you like/dislike?                                                                                                                                                                                                      \\
\hline
\multicolumn{3}{p{\textwidth}}{We are interested in creating an educational tool to support you in your reading assignments by providing you with powerful tools and reports. Here are some screenshots of the application we have developed so far \newline \textbf{Show current assignment activity screenshot} \newline \textbf{Show current assignment report screenshot}}                                                                                                                                                                                                                                                                                                                                                                                                                                                                                                                                 \\
\hline
In-Assignment Reading Support             & During a reading assignment, you can highlight, annotate, and use a dictionary to better understand a passage. We are interested in providing you with more powerful tools to support your engagement with the text. An example feature would be you can select a confusing part of the text and ask a chatbot to provide you with an explanation or summary. \newline\newline \textbf{Show the assignment screenshot} \newline\newline 2. Would this be a useful tool for you? & 2a. What other tools would be useful for you? \newline 2b. Would you like a chatbot to discuss a passage with you?                                                                                                                                                                                                                                  \\
\hline
Post-Assignment Dashboard                 & After you complete an assignment, we provide you with a dashboard that contains data reports and visualizations to help you reflect on your performance. Take a moment to look into the dashboard and try to understand what is being displayed. \newline\newline \textbf{Show the post-assignment dashboard} \newline\newline 3. Take a look at this dashboard and describe your initial thoughts. Feel free to ask us questions if something is not clear.                    & 3a. Are the visualizations easy or hard to understand? \newline 3b. What other information or diagrams would you find helpful? \newline 3c. What other kind of tools would you find helpful? \newline 3d. Would a chatbot that explains the visualizations and other information be helpful for you? \newline $\bullet$ If yes, what would you like for this chatbot to do? \newline $\bullet$ If not, why? \\
\hline
Progress \& Motivation - Student Overview & A large aspect of learning is to track your educational journey. We are interested in creating a page dedicated to helping you track your growth and learning by providing you with data about your improvements. \newline 4. How would you like this information to be presented to you?                                                                                                                                                              & 4a. What other information or diagrams would you find helpful? \newline 4b. What tools would you find helpful (Eg. highlight content topics that you might need to brush up on)?                                                                                                                                                                    \\
\hline
Final Remarks                             & 5. Is there anything else you would like to share?                                                                                                                                                                                                                                                                                                                                                                                                                                      &   \\
\bottomrule
\end{tabular}\label{tab:interview_students}
\end{table}

%% file: figures/in_depth_interview_teachers.tex
\begin{table}[]
\centering
\caption{\textbf{Teacher Interview Protocol and Questions}}
\begin{tabular}{p{2cm}|p{7cm}|p{5.5cm}}
\toprule
\textbf{Topics}                           & \textbf{Main Questions}                                                                                                                                                                                                                                                                                                                                                                                                                                                                 & \textbf{Follow-up questions / Prompts / Keywords to look for}                                                                                                                                                                                                                                                                                                              \\
\toprule
\multicolumn{3}{p{\textwidth}}{To start our conversation, we would like to ask you a couple of questions to understand your teaching goals, and objectives, and how technology can assist you}                                                                                                                                                                                                                                                                                                                                                                                                                                                                                                                                                                                                                                                                                                                                                                       \\
\hline
Role                        & 1. Can you tell us what grade and subject you teach?                                                                                                                                                                                                                                                                                                                                                                                                                       &                                                                                                                                                                                                       \\
\hline
Current Approach                        & 2. How do you currently instruct and assess students’ reading comprehension and ability?                                                                                                                                                                                                                                                                                                                                                                                                                       & 2a. What is your current approach to identifying who needs more or less reading support? \newline 2b. How do you use these insights to teach students in a personalized manner?                                                                                                                                                                                                      \\
\hline
\multicolumn{3}{p{\textwidth}}{We are interested in creating an educational tool to support you and your students in your reading assignments by providing you with powerful tools and reports to help track students’ reading comprehension and content knowledge by using eye-tracking and AI technology. Here are some screenshots of the application we have developed so far
\newline \textbf{Show current assignment auditing}
\newline \textbf{Show current assignment report}
\newline \textbf{Show current course overview}}                                                                                                                                                                                                                                                                                                                                                                                                                                                                                                                                 \\
\hline
Assignment Editing             & We want to provide you with the flexibility to create reading assignments while also considering convenience. In our application, we want to provide you a way to create assignments while leveraging automation and AI to reduce your workload, such as AI-assisted course material generation. As of now, our system supports you uploading your own passages and questions. \newline \textbf{Show the assignment auditing page} \newline\newline 3. How would you like to create reading assignments?& 3a. What would that look like, can you provide an example? \newline 3b. What are your thoughts on the use of AI to assist you? \newline 3c. What parts of assignment auditing would you like for AI to assist you (text generation, different levels of difficult versions of the assignment, question suggestions)?\\
\hline
Post-Assignment Report                 & Once your students complete an assignment, we provide you with a databoard that includes learning outcomes and process data, such as where students focused during a reading via eye-tracking. In this dashboard, we provide a way to inspect individual, group, or classwide analytics. 
\newline\textbf{Show the post-assignment dashboard}
\newline\newline
4. Take a look at this dashboard and describe your initial thoughts. Feel free to ask us questions if something is not clear.
& 
4a. Are the visualizations easy or hard to understand?
\newline 4b. What other information or diagrams would you find helpful?
\newline 4c. What other kind of tools would you find helpful?
\newline 4d. Would a chatbot that explains the visualizations and other information be helpful for you?
\newline $\bullet$ If yes, what would you like for this chatbot to do?
\newline $\bullet$ If not, why?
 \\
\hline
Course Overview & We are also interested in assisting you track your classroom’s educational journey and identify larger vital learning trends. We are interested in creating a page dedicated to visualizing your students progress and performance over a semester/course.

5. How and what information would you like to be presented to you to assist tracking your class’ progress?                                                                                                                                                             & 5a. What other information or diagrams would you find helpful? \newline 5b. What kind of tools would you find helpful (Eg. highlight struggling students)?                                                                                                                                                                    \\
\hline
Final Remarks                             & 5. Is there anything else you would like to share?                                                                                                                                                                                                                                                                                                                                                                                                                                      &   \\
\bottomrule
\end{tabular}\label{tab:interview_teachers}
\end{table}

%% file: figures/figma_workshop.tex
\begin{table}[h]
    \centering
    \caption{Protocol for Interactive Workshop}
    \begin{tabular}{p{\textwidth}}
        \toprule
        Introduction: 5 min\\
        Brief Training in How Interactive Figma Prototype Functions: 5 min \\
        Free Prototype Exploration: 10 min \\
        Course Overview: 10 min \\
        Assignment Editing (Teacher) / Assignment Activity (Student): 10 min \\
        Post-Assessment Dashboard: 20 min \\
        
        \hline
        \textbf{Course Overview} \\
        1. \textbf{Observe}: Can you describe what you see? What elements/components do you see in this page? \newline
        2. \textbf{Observe}: Would these graphs and plots help you understand your progress (score, speed, performance, reading ability)? \newline
        3. \textbf{Observe}: Rank from most important/helpful to least important helpful the visualizations you see \newline
        4. \textbf{Feedback}: What things you liked/disliked from this page? \newline
        5. \textbf{Feedback}: What other features/things you would like to have? \\\\
        \hline
        \textbf{Assignment Editing (Teacher)} \\
        1. \textbf{Observe}: Can you describe what you see? What options do you have while creating an assignment? \newline
        2. \textbf{Observe}: Would do you think about the settings for the assignment. Can you tell us which setting are useful and which are redundant? Are there more settings that you'd like? \newline
        3. \textbf{Action}: Can you tell us where will you put tags to your questions? \newline
        4. \textbf{Feedback}: What should we represent on Assignment completion page? \newline
        5. \textbf{Feedback}: Do you need any other options while editing the assignment? \\\\
        \hline
        \textbf{Assignment Activity (Student)} \\
        1. \textbf{Observe}: Can you describe what you see? What elements/components do you see in this page? \newline
        2. \textbf{Observe}: Would you find the chatbot now helpful if it you could have a conversation about the passage? Is the placement comfortable? \newline
        3. \textbf{Action}: Can you show us how you would interact (highlight, annotate, dictionary look-up, AI-request) the text? \newline
        4. \textbf{Feedback}: What things you liked/disliked from this page? \newline
        5. \textbf{Feedback}: What other features/things you would like to have? \\\\
        \hline
        \textbf{Post-Assessment Dashboard Questions} \\
        \textbf{Observe}: Can you describe what you see? What elements/components do you see in this page? \newline
        \textbf{Observe}: Can you tell us what each PDF visualization mean? How would you use them? \newline
        \textbf{Observe}: Can you rank the different visualizations of the PDF? \newline
        \textbf{Observe}: Can you tell which questions did you get wrong? \newline
        \textbf{Action}: Can you check the heatmap to tell if you read the caption on the second page? \newline
        \textbf{Action}: Can you tell us how will you find the correct answer to Q.3 ? \newline
        \textbf{Feedback}: Do you think the placement of AI chatbot is correct? \newline
        \textbf{Feedback}: What things you liked/disliked from this page? \newline
        \textbf{Feedback}: What other features/things you would like to have? \\
        \bottomrule
    \end{tabular}
    \label{tab:figma_workshop_questions}
\end{table}